# Design of a 6-bit Threshold Inverter Quantization (TIQ) Flash Analog to Digital Converter (ADC)

Noyon Kumar Sarkar[1]
noyonkumarsarkar@gmail.com
Moumita Roy[1]
moumita_m200958@ku.ac.bd
Md. Tariq Hasan[2]
mth@ece.ku.ac.bd

**Abstract**

An ADC is used to convert analog signals into binary signals. Compared with many other types of ADCs, flash converters are incredibly quick. A typical Flash ADC consists of $2^n$ resistors, $2^n-1$ op-amp comparators, and an encoder which requires more area. The resistors and comparators can be eliminated by using threshold inverter quantization (TIQ) comparators. As a voltage comparator, TIQ technique uses two cascaded CMOS inverters. So that there will be no variation in the fabrication process, and temperature. A 6-bit flash ADC based on threshold inverter quantization (TIQ) comparator was designed and software implementation was performed employing a fat tree encoder with 0.25 μm CMOS technology. The design consists of $2^n-1$ TIQ comparator arrays, a gain booster, a 1-out-of-n code generator, and a fat tree encoder. This TIQ flash ADC is simulated with the Tanner EDA Tool. Here the supply voltage is 2.5 V, input frequency of 10 kHz and 10 MHz. The power consumption of the ADC is 6.25 mW, and the propagation delay is 1.07 μs for 10 kHz input frequency. For 10 MHz input frequency, power consumption is 12.12 mW and propagation delay is 947.14 ms.

**Keywords**: threshold inverter quantization, comparator, gain booster, 1-out-of-n code generator, fat tree encoder.

[1]Electronics and Communication Engineering Discipline, Khulna University, Bangladesh.
[2]Associate Professor, Electronics and Communication Engineering Discipline, Khulna University, Bangladesh.

# Design of a 6-bit Threshold Inverter Quantization (TIQ) Flash Analog to Digital Converter (ADC)

**Introduction**

The system that transforms analog events into a digital signal that can be processed and stored on computers is known as an analog to digital converter or ADC. Modern electronic devices use digital signals to operate, but since signals in the actual world take an analog form, it is necessary to convert analog impulses to digital signals. The advantages of digital circuits over analog circuits are larger because they are faster and transmit data with less quality loss while being more accurate. As a result, we need ADC to transform analog signals into digital ones. There are several ADC architectures. All but the most specialized ADCs are constructed as integrated circuits because of the complexity and requirement for perfectly matched components. Flash is the top option among many ADCs since it provides fast speed.

Challenges of Designing ADC

To make an ADC there are some challenges because ADC should be adaptable for SoC implementation. The main challenges in designing ADC for the SoC are low voltage, low power, and high speed. In mixed-signal ICs, one of the toughest problems is low-voltage functioning. The supply voltage is decreased to 2.5V by downscaling the minimum channel length to 0.25 μm in this project. Low power consumption is the next difficulty. It is an important concern for portable devices. When it comes to high speed, the ADCs made with the CMOS process in this project. It also reduces space.

**Motivation**

Without ADC modern technology can't be imagined. The applications of ADC are boundless. Some important applications of ADC are:

- ❒ In Digital cameras images and videos are shown and stored in digital form. But in real life when images and videos are captured, it remains in analog form. Here, ADC converts the analog data into digital form so that we can see images on the screen.
- ❒ In the medical sector, x-ray and MRI also uses ADC to convert images into digital form.
- ❒ ADC is used in digital oscilloscopes for converting analog signal into digital signal.

- ❐ Naturally any kind of sound is in analog form. If we want to transmit or store these sounds, we need to convert those into digital form. Digital voice signals are used for the operation of cell phones. ADC converts analog voice into digital voice before sending it to a cell phone transmitter.
- ❐ A huge number of sensors are used for different purposes. Like temperature sensors, water sensors, gas sensors, humidity sensors etc. ADC is used in all these sensors to measure the value and shown in display. Because usually all these sensor's data are in analog form.

The huge uses of ADC motivate us to design an ADC. So, we decided to design a TIQ flash ADC. Because TIQ is suitable for SoC implementation. There are many types of ADC in which flash is very fast and its maximum resolution is 12 bits. For this reason, we choose flash to design TIQ flash ADC.

**Challenges of Designing ADC**

To make an ADC there are some challenges because ADC should be adaptable for SoC implementation. Designing ADCs for a complete System-on-Chip (SoC) involves addressing key challenges such as low voltage, low power consumption, and high speed.

In mixed-signal ICs, one of the toughest problems is low voltage functioning. The power supply voltage is decreased to 2.5 V by downscaling the minimum channel length to 0.25 μm. However, the SIA Roadmap's estimated minimum supply voltage for analog circuits does not adhere to the reduction in digital supply voltage. The task of creating an ADC that operates at low voltage is challenging for mixed-signal circuit designers, mainly due to the relatively high threshold voltage of transistors. As a result, an ADC should be operated within a limited voltage range.

Low power consumption is the next difficulty. One of the key concerns in the market for portable devices is power consumption reduction. For portable devices, ADCs and digital circuits should be combined on a single chip. In order to increase battery life, low power techniques are now being incorporated into the design of all battery-powered products. Similar to this, ADCs require a low power architecture or approach.

When it comes to high speed, the ADCs made with an advanced BiCMOS process have a sampling speed of about 200 mega samples per second (MSPS). One of the leading ADC foundries has developed recent high-speed ADCs using a bipolar process, which can operate at up to 1.5 Giga samples per second (GSPS). These ADCs are suitable for use in various applications, such as digital oscilloscopes, digital RF/IF signal processing, direct RF down-conversion, and radar/ECM systems This Maxim Integrated Products ADC uses bipolar solid-state technology and is fast enough for SoC implementation, but it costs too much. The entire SoC has two limitations: cost and high-speed operation. Consequently, to reduce the speed disparity between a CPU and an ADC in the entire SoC implementation, it is necessary for an ADC design to be rapid and also cost-effective.

**Solid-State Technology**

The rate at which an ADC operates can also be impacted by the type of solid-state technology employed to construct the converter. CMOS, bipolar, and gallium arsenide (GaAs) technologies are the three types of solid-state technologies now employed for high-speed ADC implementations. CMOS technology is the slowest and GaAs technology is the fastest of the three. Bipolar technology is compatible with CMOS technology and enables speedier operation. In contrast to conventional CMOS technology, BiCMOS takes more processing steps and is more expensive. Although GaAs performs better, it is not compatible with the current CMOS technology, which is widely utilized in consumer electronics. SoC devices are best served by mixed-signal circuit implementation using simply conventional CMOS technology. Therefore, we suggest a low-power, high-speed CMOS flash design that uses the Threshold Inverter Quantization (TIQ) method. Using the conventional CMOS logic circuits chosen for SoC implementation, the TIQ approach boosts ADC speed. A simplified comparator design is the key benefit of the TIQ based CMOS flash ADC design. This is achieved by using digital inverters as analog voltage comparators, which are faster and less complex than high-gain differential input voltage comparators. Reference voltages are also no longer required because the TIQ flash ADC does away with the necessity for a resistive network circuit. High speed and low power consumption are both possible due to the comparator's simplicity. This highlighted ADC is an ideal fit for a complete SoC implementation, as it can be built using standard CMOS logic technology. However, it should be noted that the single-ended inverter comparator used in this design is more susceptible to noise and may exhibit variations in the ADC input range due to differences in process parameters during fabrication. For a TIQ flash ADC to be successfully implemented, these two requirements must be properly taken into account.

**Literature Review**

Without ADC modern technology can't be imagined. A TIQ flash ADC is a suitable option for converting signals with high speed and high accuracy because of its ability to convert signals quickly, consume less power, achieve high precision, adapt to various signal amplitudes and frequencies, and ensure reliable performance.

The development of flash ADC has been a major focus of research during this era, with many significant contributions aimed at improving its performance and capabilities. A research study in 2008 proposed a resolution of adaptive TIQ ADCs, aimed to achieve high conversion speed and to reduce the analog nodes in the ADC. The designed ADC was able to operate at different precisions (3-bit, 4-bit, 5-bit, and 6-bit) depending on control inputs, making it a true variable resolution ADC. A 3.3V power supply voltage and AMS 0.35 μm CMOS technology were used to implement the design.

Parvaiz and Mir designed a 4-bit flash ADC using TIQ and a TC-to-BC encoder with reduced power consumption in 2013. The system achieved a high sampling frequency and a low power demand, using 0.12 μm technology with the ADS 2006A tool. Power consumption of that design was 4.25 mW. They varied the reference voltage from 0.58 V to 1.21 V with 20 MHz sinusoidal input frequency.

In 2017, Talukder and Sarker proposed a 3-bit TIQ comparator-based Flash ADC design technique with a mux-based encoding strategy to improve conversion speed. The ADC was implemented using Cadence Virtuoso in CMOS 0.5 μm technology. The power supply voltage of that design was 5V. The lowest and highest switching voltage was 1.0 V and 3.70 V respectively. They also designed a layout of the ADC chip. The layout's size was determined to be 114 μm × 140.1 μm using 0.50 μm technology.

In 2021, a 5-bit flash ADC was developed using 0.18 μm CMOS technology which was thermally reliable. In the design of the ADC, the TIQ comparator demonstrated a minimal shift of only 9 mV in its switching voltage across a temperature range of -20º to 120º C. They also used a multiplexer-based encoder.

**Methodology**

The proposed flash ADC employs the threshold inverter quantization (TIQ) technique, which utilizes conventional CMOS technology and is suitable for microprocessor fabrication. This technique enables the flash ADC to operate at high speeds and with low power consumption. The name of the technique comes from the use of two cascading inverters as a voltage comparator. The transistor length and width ratio ($Z = L/W$) of the inverters determine the internal reference voltages that the voltage comparators used to compare the input voltage. Therefore, we do not require the series resistor network that is utilized in a traditional flash ADC. Gain boosters are used to sharpen the comparator output and provide a full digital output voltage swing. In the flash ADC, the comparator outputs are transformed into a thermometer code and a binary code through a 1-out-of-*n* code generator and an encoder, as depicted in Figure. These two phases of transformation enable the conversion of the analog input signal into a digital output.

**Typical Flash ADC**

Typical Flash ADC consists of a resistive network, op–amp comparator and encoder. An example of a flash ADC is shown in Figure 1 $2^n$-1 comparators are required for an *n*-bit ADC. A reference voltage is provided by an external reference source for each comparator using resistor.

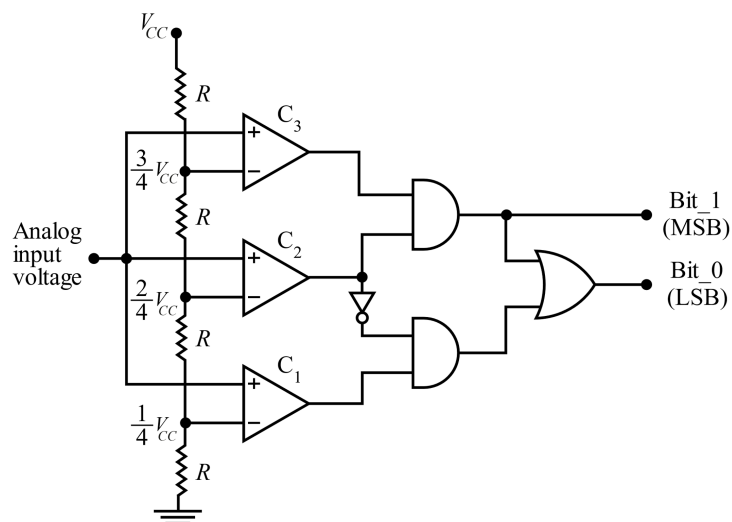

Figure 1 : Typical Flash ADC.

In Figure 1 from the biggest reference voltage $\frac{3}{4}V$ to the smallest reference voltage $\frac{1}{4}V$ These values are equally spaced by $V_{LSB}$. Each comparator has a connection to an analog input, allowing each comparator output to be generated in a single cycle. The encoder converts the thermometer code, which is the set of comparators' digital output, into a binary code.

**Block Diagram of Designed TIQ Flash ADC**

For high speed and low power consumption, the TIQ comparator uses two cascading CMOS inverters as a comparator. By comparing a reference voltage ($V_{ref}$) with the input voltage

($V_{in}$), it transforms the input voltage ($V_{in}$) into a logic "1" or "0". The comparator outputs "1" if $V_{in}$ exceeds $V_{ref}$ and "0" if the input voltage is less than the reference voltage. The fully differential latch comparator and dynamic comparator are often utilized comparator topologies in CMOS ADC design. The former is occasionally referred to as a "clocked comparator". whereas the latter is frequently referred to as a "auto-zero comparator" or "chopper comparator". Such comparators often use bipolar transistor technology to achieve high speed. To implement an SoC that combines a high-speed ADC and a digital signal processor on the same substrate, BiCMOS technology would be necessary. In order to achieve both high speed and low power consumption, the TIQ comparator utilizes two CMOS inverters arranged in a cascading manner as the comparator. In order to create a high-speed flash ADC, Tangel used this TIQ comparator. TIQ comparator eliminates resistor ladder and op-amp so that it minimizes the size of the ADC. To complete the design gain booster, 1-out-of-n code generator, and encoder are also required. Figure 2 shows the block diagram of 6-bit TIQ flash ADC.

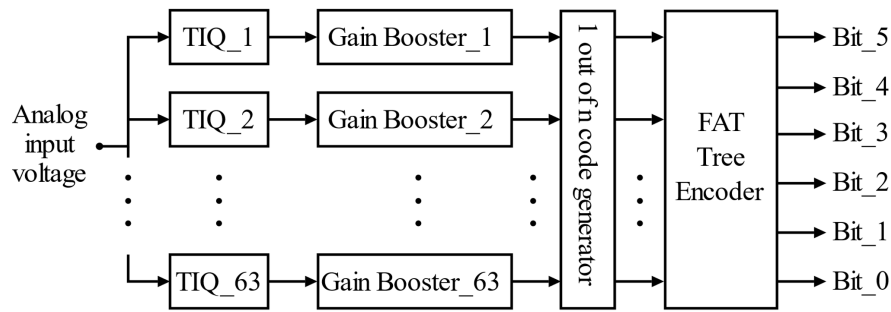

Figure 2: Block diagram of 6-bit TIQ ADC.

**TIQ Comparator Design**

The $V_{in} = V_{out} = \frac{V_{DD}}{2}$ in the VTC of an inverter is referred to as the inverter threshold ($V_{ref}$). Mathematically,

$$V_{ref} = \frac{r(V_{DD} - |V_{Tp}|) + V_{Tn}}{1+r} \qquad (1)$$

$$\text{Where,}\; r = \sqrt{\frac{k_p}{k_n}} = \sqrt{\frac{\mu_p W_p}{\mu_n W_n}} \qquad (2)$$

where $V_{Tn}$ and $V_{Tp}$ stand for the respective threshold voltages of n-MOS and p-MOS devices. Figure 3 shows the VTC curve of the CMOS inverter. According to the W/L ratios of p-MOS and n-MOS, $V_{ref}$ controls the analog input signal quantization level at the first inverter. But the length of both transistors are kept constant. To boost voltage gain, a second inverter is used.

From Equation,

$$V_{ref} = \frac{\sqrt{\frac{\mu_p W_p}{\mu_n W_n}}(V_{DD} - |V_{Tp}|) + V_{Tn}}{1 + \sqrt{\frac{\mu_p W_p}{\mu_n W_n}}} \quad (3)$$

By changing transistor sizes ($W_p/W_n$) the reference voltage $V_{ref}$ can be changed. Increasing $W_p$ makes $V_{ref}$ bigger and increasing $W_n$ makes $V_{ref}$ smaller.

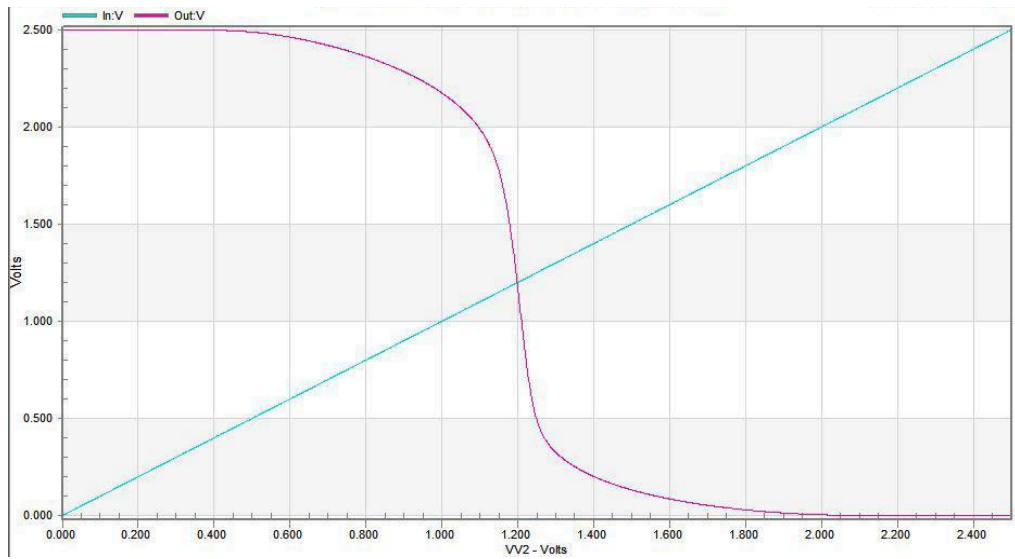

Figure 3: VTC curve of CMOS.

In the TIQ comparator by changing transistor size, switching voltage can be changed. It means in TIQ flash ADC, reference voltages ($V_{ref}$) are internally built-in by sizing transistors. Figure 4 shows the circuit diagram of TIQ.

**TIQ Comparator Designing Algorithm**

To design $n$-bit ADC, $2^n-1$ reference voltages are needed. To find these voltage,

At first maximum and minimum reference voltage is needed to be calculated.

Minimum reference voltage, $V_L = V_{ref} - \frac{V_{DD}}{2|A_v|}$

Maximum reference voltage, $V_H = V_{ref} + \frac{V_{DD}}{2|A_v|}$

CMOS inverters have a channel length that is as short as possible (to minimize the area and maximum the density). The output resistances are relatively small and a typical value of $\frac{V_{out}}{V_{in}}$ is - 5 to – 10. The next step is to calculate the voltage step $V_{LSB}$. The equation of calculating

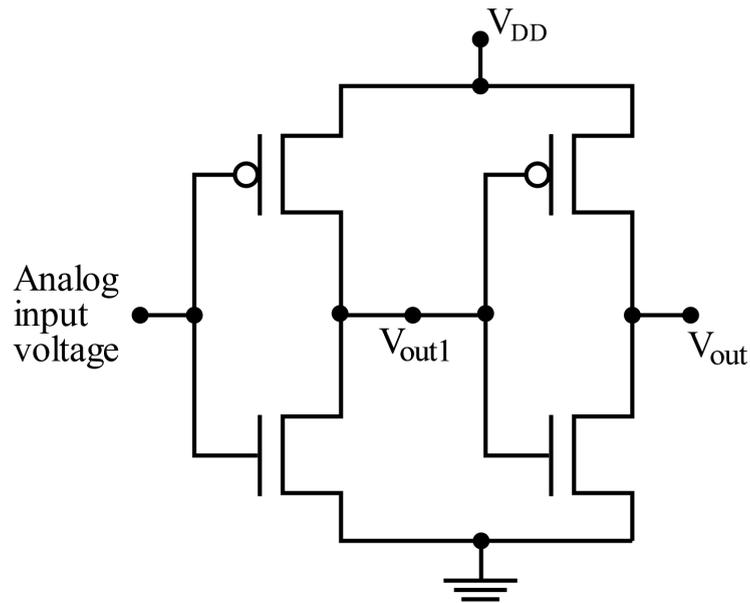

Figure 4: Circuit diagram of a TIQ.

$V_{LSB}$ is $V_{LSB} = \frac{V_H - V_L}{2^n - 2}$. After that it is required to find $2^n$-1 ideally quantized $V_{ref}$ voltages from $V_L$ through $V_H$ equally spaced by $V_{LSB}$ for linear quantization. Then we select the actual $2^n$-1 $V_{ref}$ close to the ideal $V_{ref}$ from the calculated $V_{ref}$ except the two outer $V_{ref}$. For 6-bit TIQ Flash ADC we designed $2^6 - 1$ comparators. Figure 5 shows the VTC curve of 63 comparators which are designed for 6-bit TIQ flash ADC.

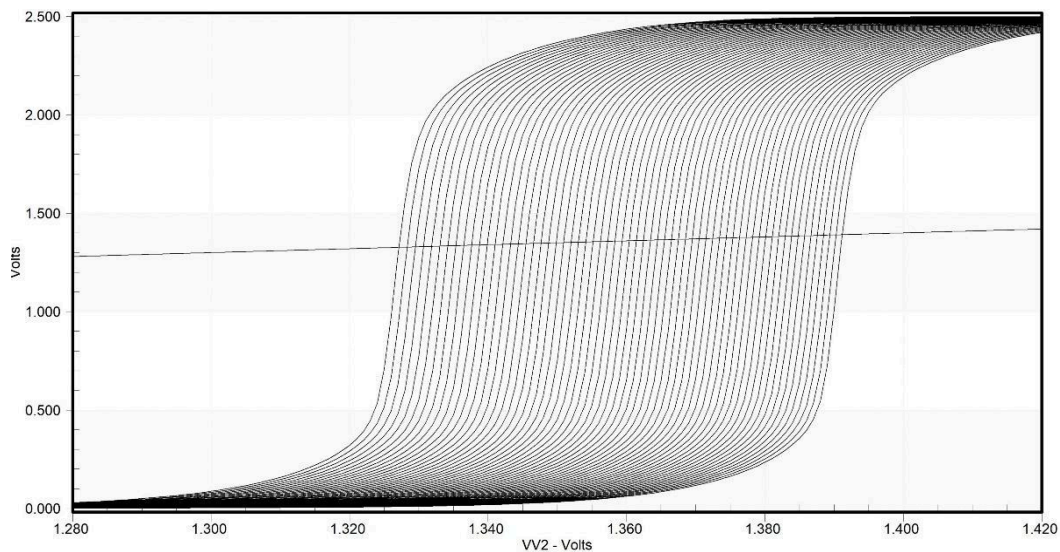

Figure 5: The VTC of the 6-bit TIQ comparator.

**Gain Booster**

Gain Booster is used for sharper output. The design of the gain booster is the same as TIQ. But the length and width of every gain booster are small and identical. It provides full digital output voltage swing. The circuit for the comparator is shared by each gain booster, which includes two cascading inverters, but the size of transistors are all identically small. A gain booster is used to augment a comparator's output voltage gain so that it can provide a full digital output voltage swing. Figures 6(a) and 6(b), respectively, display the gain booster's propagation delay and voltage gain as a function of transistor length variation.

The voltage gain, however, has a logarithmic trend whereas the propagation delay has a trend that is practically caused by increasing the transistor length. Therefore, when determining the size of the gain booster, voltage gain and propagation delay should always be taken into account simultaneously.

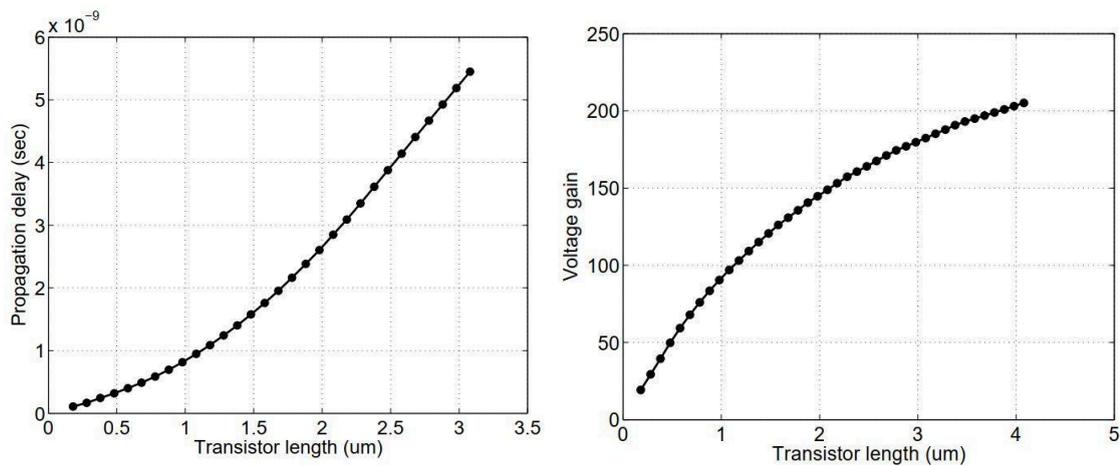

Figure 6 : (a)Propagation Delay of Gain Booster, (b) Voltage Gain of Gain Booster.

**1-out-of-*n* code generator**

A thermometer code is produced by a TIQ comparator. In two processes, an encoder is used to translate the thermometer code into binary codes. Using a 1-out-of-*n* code generator, the thermometer code is transformed into 1-out-of-*n* code. An AND gate and a NOT gate is required to make a 1-out-of-*n* code generator for one thermometer code. Figure 7 shows the design of the 1-out-of-n code generator. And Figure 6 shows the schematic diagram designed in tanner EDA tools.

**Fat tree encoder**

The main advantage of the fat tree encoder over the other encoders is its high encoding speed and low power consumption. The fat tree of OR gates is an all-digital circuit and it does not require a clock signal. The fat tree circuit signal complexity is $O(log_2N)$. We propose the use of a fat tree encoder to improve the speed of the encoder, as it is typically the bottleneck in the speed of a flash ADC. The fat tree encoder's rapid encoding speed and low power usage give it an edge over the other encoders. A 3-bit fat tree encoder example is shown in Figure 8. The leaf nodes of the tree (from a6 to a0) display the 1-out-of-8 code, that is the outcome of the "01" generator. The tree's root is where the 3-bit binary output (Bit_2; Bit_1; Bit_0) is found.

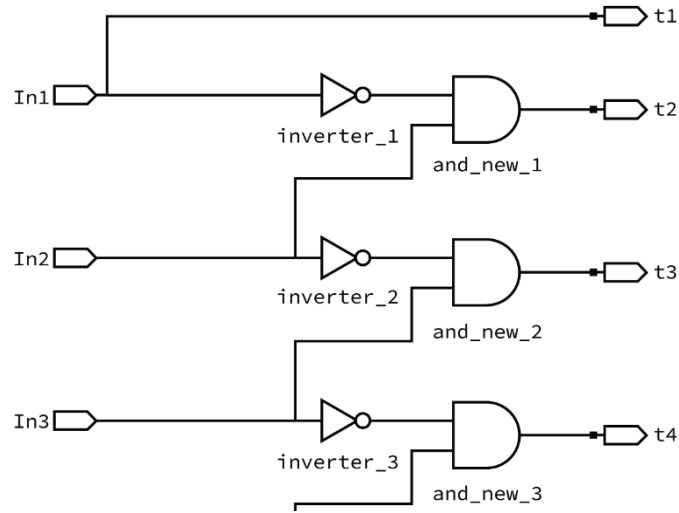

Figure 7: The schematic diagram of the 1-out-of-n code generator.

The flash ADC output is obtained by performing a logical OR operation on the output signals of the leaf nodes, which depends on the truth table shown in Table 1. To achieve this, the output signals from the six 2-input OR gates are sent to the parent nodes, and finally, the three output signals (Bit_2, Bit_1, and Bit_0) are obtained by performing OR operations on the signals at the top level of the tree. From leaf to root, there are more edges as depicted by the tree. The fat tree encoder is the term given to this new form of encoder. The fat tree encoder's tree-based architecture results in a signal latency of $O(log_2N)$. As a result, it is considerably faster than ROM type encoders; for instance, in the case of 3-bit encoding, there are only 2 OR gate delays. To improve the efficiency of the implementation, it is possible to convert the multiple OR gates to NOR or NAND gates instead. This can reduce the number of gates required and simplify the circuit design. Additionally, the fat tree encoder is simple to pipeline at any level of the tree. While the ROM-type encoder has the advantage of being faster than other types of ADCs, it is also much more complex to design and automate. This presents a challenge for improving the implementation of the TIQ flash ADC, particularly with regards to automating the design of the 3-dimensional fat tree encoder.

$$Bit\_2 = (a6+a5) + (a4+a3)$$
$$Bit\_1 = (a6+a5) + (a1+a2)$$
$$Bit\_0 = (a6+a4) + (a0+a2)$$

The main focus of this design is to generate a 63 TIQ comparator, gain booster, 1-out of-64 code generator and fat tree encoder. In this chapter we discussed how to design a TIQ comparator, gain booster, 1-out-of-$n$ code generator, and fat tree encoder. In the next chapter we will show our simulation result.

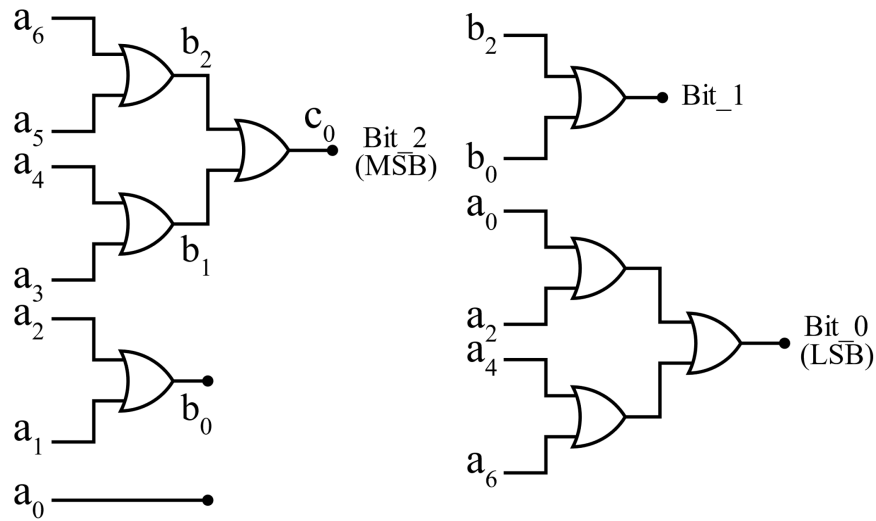

Figure 8: 3-bit Fat Tree Encoder.

Table 1 :Output of leaf nodes.

| bit (2:0) | 000 | 001 | 010 | 011 | 100 | 101 | 110 | 111 |
|---|---|---|---|---|---|---|---|---|
| a6 | 0 | 0 | 0 | 0 | 0 | 0 | 0 | 1 |
| a5 | 0 | 0 | 0 | 0 | 0 | 0 | 1 | 0 |
| a4 | 0 | 0 | 0 | 0 | 0 | 1 | 0 | 0 |
| a3 | 0 | 0 | 0 | 0 | 1 | 0 | 0 | 0 |
| a2 | 0 | 0 | 0 | 1 | 0 | 0 | 0 | 0 |
| a1 | 0 | 0 | 1 | 0 | 0 | 0 | 0 | 0 |
| a0 | 0 | 1 | 0 | 0 | 0 | 0 | 0 | 0 |
| a7 | 0 | 0 | 0 | 0 | 0 | 0 | 0 | 0 |

**Simulation Results**

Tanner EDA tool is used to design and simulate the 6-bit TIQ flash ADC. Propagation delay, consumption, and noise are measured in this simulation. Figure 9 shows the output of the designed ADC at 10 kHz input frequency. At room temperature, the propagation delay for a 10 kHz input frequency is 1.070 µs. Table 2 shows the propagation delay at different temperatures. In Figure 10, inoise shows the input noise spectral density and onoise shows the output noise spectral density.

Table 2: Propagation delays of the 6-bit TIQ flash ADC in different temperatures.

| °C | -40 | -30 | -20 | -10 | 0 | 10 | 25 | 40 | 55 | 70 | 80 | 90 | 100 |
|---|---|---|---|---|---|---|---|---|---|---|---|---|---|
| Propagation Delay (μs) | 1.281 | 1.239 | 1.200 | 1.165 | 1.350 | 1.107 | 1.070 | 1.039 | 1.014 | 0.993 | 0.981 | 0.971 | 0.962 |

Table 3, shows the power consumption at 10 kHz analog input frequency.

Table 3: Power consumption of 6-bit TIQ flash ADC.

| Average Power Consumed (mW) | 6.253492 |
|---|---|
| Maximum Power Consumed (mW) | 96.5834 |
| Minimum Power Consumed (μW) | 0.07483 |

Table 4, shows the average power consumption at different temperatures.

Table 4: Average power consumption at different temperatures.

| °C | -30° | -20° | -10° | 0° | 10° | 25° | 40° | 55° | 70° |
|---|---|---|---|---|---|---|---|---|---|
| Power Consumed (mW) | 7.709 | 7.415 | 7.135 | 6.860 | 6.253 | 6.253 | 5.917 | 5.619 | 5.314 |

Table 5, shows the summary of the simulation of 6-bit TIQ Flash ADC in 0.25 μm technology.

Table 5: Summary of simulation at room temperature.

| Resolution | 6-bit | |
|---|---|---|
| CMOS Technology | 0.25 μm | |
| Total MOSFET | 2188 | |
| Power Supply | 2.5 V | |
| $V_{FSR}$ | 1.3349V – 1.3995V | |
| $V_{LSB}$ | 1 mV | |
| Simulation Frequency | 10 kHz | 10 MHz |
| Power Consumption | 6.25 mW | 10.09 mW |
| Propagation Delay | 1.070 μs | 947.14 ms |

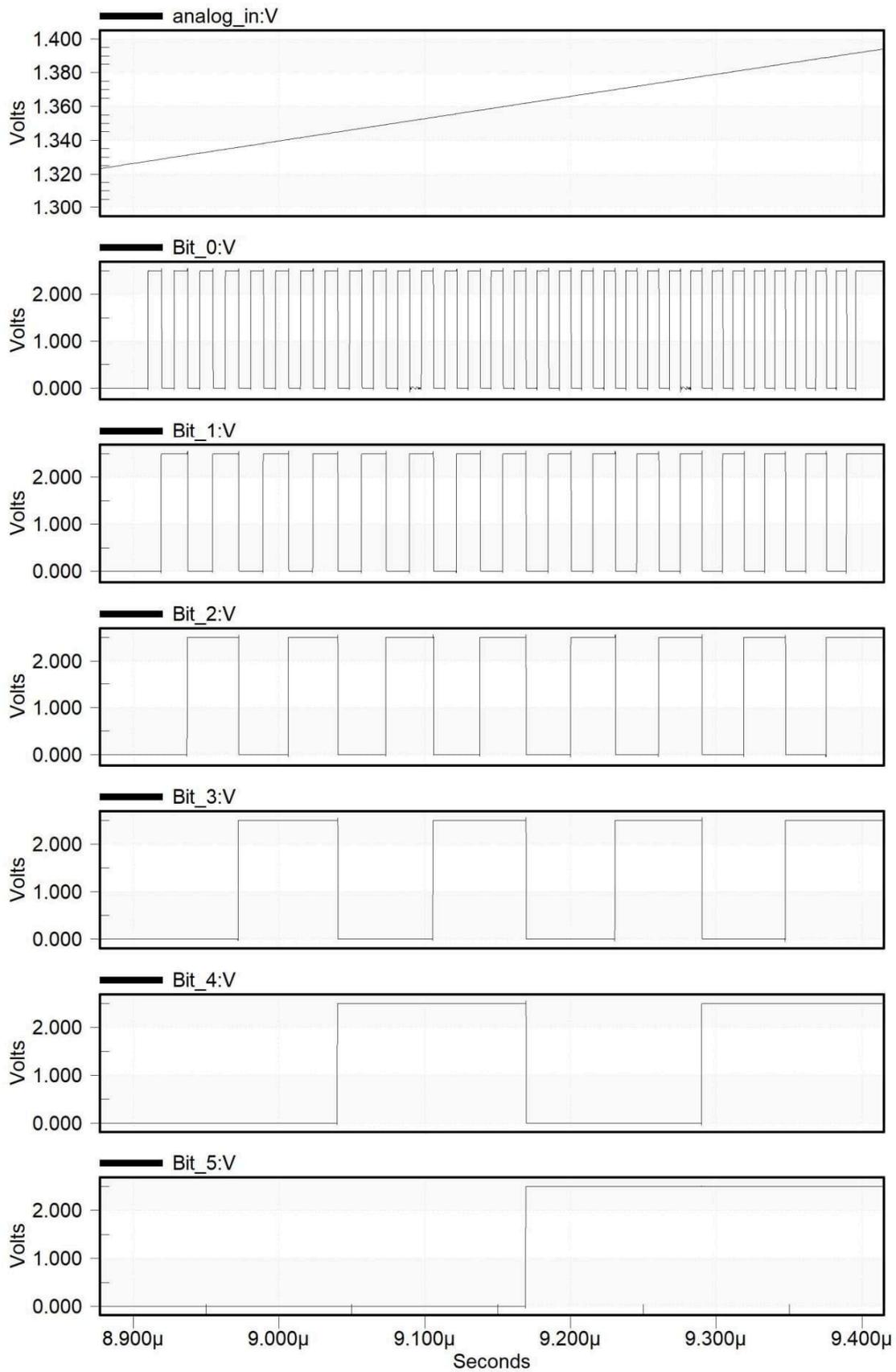

Figure 9: Input and output waveforms of the 6-bit TIQ flash ADC.

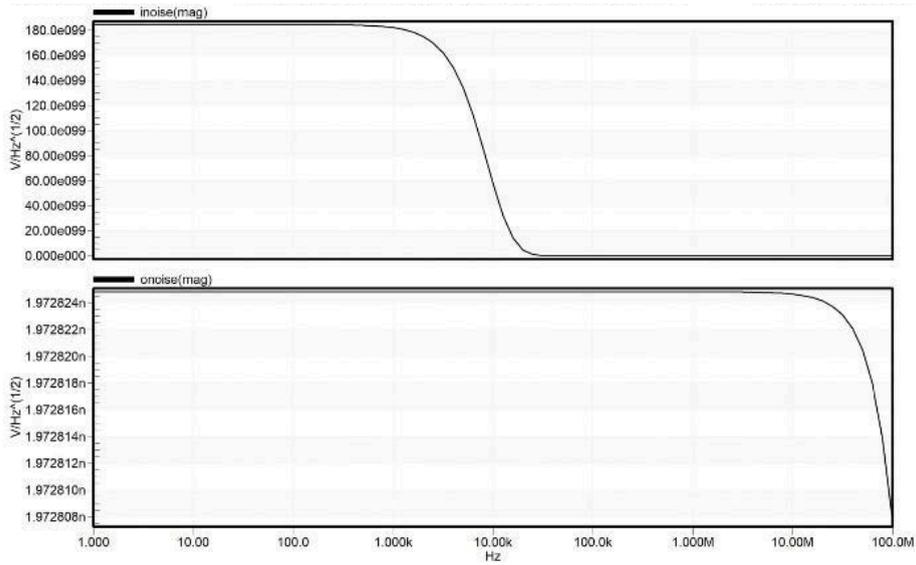

Figure 10: Input and output noise spectral density magnitude.

Figure 11, shows the propagation delay at different temperatures.

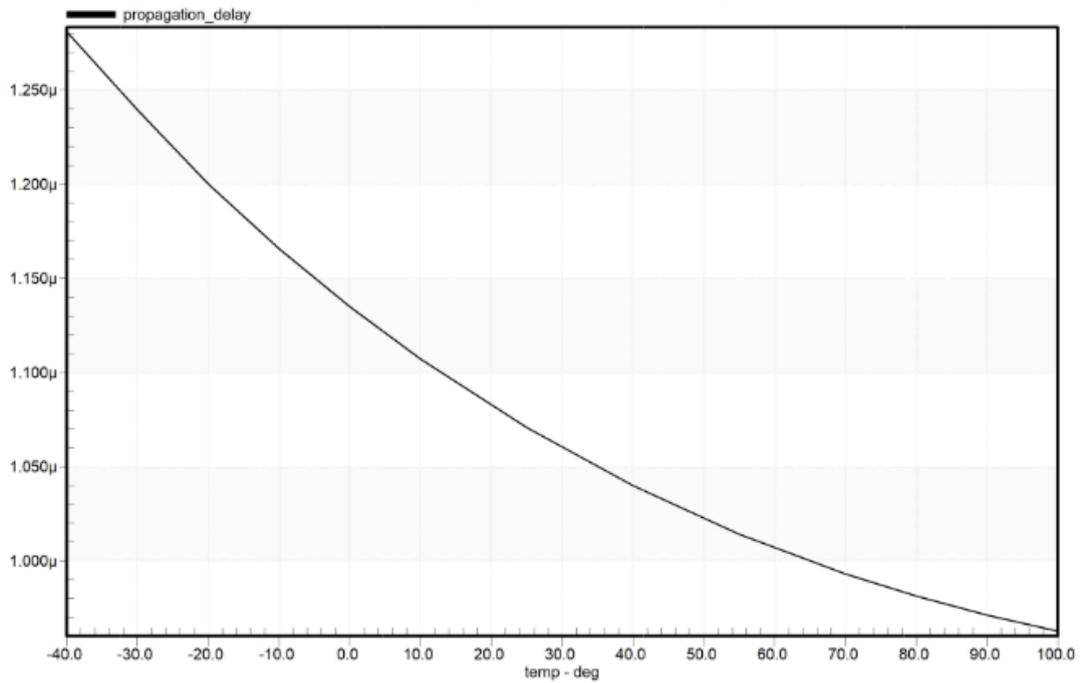

Figure 11: The propagation delay at different temperatures.

**Conclusion**

The TIQ flash ADC has been designed with 0.25 μm CMOS technology for the low voltage operation to confirm the functionality at a low power supply voltage. The TIQ flash ADC operated correctly at high speeds, with minimal linearity problems without any additional circuitry and efficient analog-to-digital conversion technique widely used in various applications, including wireless communications, radar, and computer network adaptors. Its key advantages include high-speed conversion, low power consumption, and simplicity in design. However, TIQ flash ADC design requires careful consideration of various parameters, such as comparator sizing, $V_{ref}$ selection, and encoder design. In addition, noise analysis is critical for determining the impact of noise on the ADC's performance. Despite the challenges, the TIQ flash ADC remains a popular choice for high-speed analog-to-digital conversion due to its fast conversion speed and reliable performance.

Since an inverter is often quick and consumes less energy at the front-end of the ADC, the TIQ flash ADC's key advantages are its high speed and low energy consumption using CMOS technology as standard.